\documentclass[aps, prl, reprint, longbibliography, 
superscriptaddress]{revtex4-1}

\usepackage{hyperref}
\usepackage{graphicx}
\usepackage[utf8]{inputenc}
\usepackage{xcolor}
\usepackage{amsmath, amssymb}
\usepackage{physics}

\begin{document}

\preprint{APS/123-QED}

\title{ Supersolid Striped Droplets in a Raman 
Spin-Orbit-Coupled System}

\author{J. S\'anchez-Baena}
\email{juan.sanchez.baena@upc.edu}
 \author{J. Boronat}
 \email{jordi.boronat@upc.edu}
 \author{F. Mazzanti}
 \email{ferran.mazzanti@upc.edu}%
\affiliation{%
 Departament de F\'isica, Universitat Polit\`ecnica de 
Catalunya,
Campus Nord B4-B5, E-08034, Barcelona, Spain\\
}%

\date{\today}

\begin{abstract}

We analyze the role played by quantum fluctuations on a Raman Spin-Orbit Coupled system in the stripe phase. We show that beyond mean-field effects stabilize the collapse predicted by mean-field theory and induce the emergence of two phases: a gas and a liquid, which also show spatial periodicity along a privileged direction. We show that the energetically favored phase is determined by the Raman coupling and the spin-dependent scattering lengths. We obtain the ground-state solution of the finite system by solving the extended Gross-Pitaevskii equation and find self-bound, droplet-like solutions that feature internal structure through a striped pattern. We estimate the critical number for binding associated to these droplets and show that their value is experimentally accessible. We report an approximate energy functional in order to ease the evaluation of the Lee-Huang-Yang correction in practical terms.

\end{abstract}

\maketitle

Spin-Orbit Coupling (SOC), which denotes the interplay between a
particle's momentum and its spin, has been a subject of interest in
the recent years, both theoretically and experimentally. It plays an important role in a wide variety of exotic quantum phenomena, such as  topological insulators~\cite{hasan10} and
topological superconductors~\cite{sato17}. SOC is a relativistic effect naturally found in electronic and atomic systems. However, it can also be synthetically
engineered~\cite{zhang18} in ultracold atomic gases, which represent an excellent platform to study the physics of SOC due to their high controllability and tunability.  In particular, Raman SOC was first implemented experimentally by inducing a Raman
coupling via two laser beams on an atomic Lambda type
configuration~\cite{spielman11,li15}. Raman SOC has been realized with $^{87}$Rb atoms, both in the
continuum~\cite{spielman11} and in a lattice~\cite{hamner15,bersano19}, and with other atomic species such as $^6$Li~\cite{cheuk12} and $^{40}$K~\cite{wang12}. In this context, two hyperfine states of an atom are labeled as pseudospin states.

In this Letter, we focus in Raman SOC, which couples the 
linear momentum of an atom with its spin according to 
\begin{equation}
 \hat{W}^{\text{SOC}} =
\frac{ \hbar k_0}{m} \hat{P}_x \hat{\sigma}_z +
\frac{\hbar^2 k_0^2}{2m} - \frac{\Omega}{2}
\hat{\sigma}_x   \ , 
\label{Wraman}
\end{equation}
with $m$ the atomic mass, $\hat{P}_x$ the $x$-component of
the momentum operator, $\hat\sigma_x$ and $\hat\sigma_z$ 
the Pauli matrices, $\Omega$ the
Raman coupling, and $k_0$ the magnitude of the wave vector
difference between the two laser beams.

We are particularly interested in the emergence of a stripe phase, which arises from the breaking of two symmetries: a gauge symmetry, giving
rise to off-diagonal long-range order, and spatial symmetry, seen as a
periodic density modulation in space~\cite{li13}. In contrast to other systems featuring spatial ordering like dipoles~\cite{macia14,bombin17}, this property is present in SOC systems even in ultradilute conditions. The stripe phase has  been both predicted theoretically~\cite{li13,li15} and detected experimentally~\cite{ketterle17}, and the resulting stripes have been shown to be superfluid~\cite{li13,chen18,sanchez20}, therefore being often referred as supersolid stripes or superstripes. It has also been shown that the increase of interatomic correlations enhances the domain of the stripe phase in the phase diagram of Raman SOC systems~\cite{sanchez20}.

As it happens in ultradilute Bose-Bose mixtures, systems featuring a spin-dependent interaction can become unstable at the mean-field level for some  values of the spin-dependent scattering lengths. A well known result for unstable Bose-Bose mixtures is that quantum fluctuations can stabilize the system through the Lee-Huang-Yang (LHY) energy correction, giving rise to liquid droplets~\cite{petrov15}. In this Letter, we investigate if the same mechanism also holds in a mixture under Raman SOC in the stripe phase. While in previous works, beyond mean-field properties like the excitation spectrum and the static structure factor of the stripe phase have been reported~\cite{li13}, the complete LHY energy correction containing SOC terms has not been derived. We evaluate this term in order to estimate the role played by quantum fluctuations in a stripe state under Raman SOC which is unstable at the mean-field level.

We study an $N$-particle system governed by the Hamiltonian:
\begin{equation}
\hat{H} = \sum_{i} \left[ \frac{\hat P_i^2}{2m} + \hat{W}_i^{\text{SOC}} \right] +
\sum_{i<j} \hat{V}_{ij} = \sum_{i} \hat{H}_{0,i} + \sum_{i<j} \hat{V}_{ij} \ ,
\label{Hamiltonian}
\end{equation}
with $\hat{V}_{ij}$ a two-body interaction given by
\begin{equation}
\hat{V}_{ij} = \sum_{s_1,s_2} \frac{4 \pi \hbar^2 a_{s_1,s_2}}{m} \delta(\vec{r}_i-\vec{r}_j) \ket{s_1,s_2}\bra{s_1,s_2} \ ,
\label{pot_mf}
\end{equation}
where $s_i$ is the spin coordinate of the $i$-th particle and $a_{s_1,s_2}$ are the spin-dependent scattering lengths. To reduce the number of variables, we set $a \equiv a_{+1,+1} = a_{-1,-1} \neq a_{+1,-1} = a_{-1,+1}$. We introduce the parameter $\gamma = ( a - a_{+1,-1} )/( a + a_{+1,-1} )$, which indicates the contrast of the spin-dependent interaction. Results are given in reduced units, with length and energy scales given by $a_0 =1 /k_0 $ and $\epsilon_0 = \hbar^2 k_0^2 / 2 m $, respectively.

In second quantization, the system can be described by the field operator spinor, $\hat{\Psi}(\vec{r}) = (\hat{\psi}^{+1}(\vec{r}) \text{ } \hat{\psi}^{-1}(\vec{r}))^\tau$,  with the $\pm 1$ indexes indicating the spin component. Within the mean-field approximation, the dynamics of the system are driven by the time-dependent Gross-Pitaevskii 
equation~\cite{martone12},
\begin{equation}
 i \hbar \dv{\hat{\Psi}(t)}{t} = \hat{H}_0 \hat{\Psi} + \left[ \frac{2 G_1}{n} \hat{\Psi}^{\dagger} \hat{\Psi} + \frac{2 G_2}{n} \left( \hat{\Psi}^{\dagger} \hat{\sigma}_z \hat{\Psi} \right) \hat{\sigma}_z \right] \hat{\Psi} \text{ .}
 \label{tgpe}
\end{equation}
with $G_1 = n(g_{+1,+1}+g_{+1,-1})/4$ and $G_2 = n(g_{+1,+1}-g_{+1,-1})/4$. In the stripe phase, the condensate wave function can be written as~\cite{li13,martone18}
\begin{equation}
  \vec{\psi}_0(\vec{r}) = \frac{1}{\sqrt{V}} \sum_{n \in \mathbb{Z}} \vec{\psi}_{0,n} e^{i k_1 x + 2 i n k_1 x}
  \label{stripe_cond}
\end{equation}
with the amplitudes fulfilling the condition $\psi^{\pm 1}_{0,j} = (\psi^{\mp 1}_{0,-j-1})^*$~\cite{martone18}. The value of the amplitudes $\vec{\psi}_{0,n}$ and the stripe momentum $k_1$ can be obtained by minimizing the mean-field energy. We do that using the Simulated Annealing algorithm~\cite{harland88}. The effect of quantum fluctuations can be included following the Bogoliubov scheme. Within this formalism, the time-dependent field operator is written as
\begin{equation}
  \hat{\Psi}(t) = e^{-i \mu t/\hbar} \left( \hat{\psi}_0 + \hat{\delta \Psi} (t)  \right)
 \label{tdpsi_soc}
\end{equation}
where $\hat{\psi}_0 = \psi_0(\vec{r}) \hat{a}_0$ corresponds to the condensate state and $\hat{\delta \Psi} (t)$ accounting for the quantum fluctuations. For the stripe phase, the quantum fluctuations operator can be decomposed as~\cite{martone18}:
\begin{align}
 \delta \hat{\Psi}\left( \vec{r},t \right) &= \sum_{\substack{0< k_x < k_1\\ 0 < k_y,k_z < \infty \\ l}} \vec{f}_{\vec{k_1} + \vec{k},l}(\vec{k},\vec{r}) e^{-i E_{\vec{k_1} + \vec{k},l} t/\hbar} \hat{b}_{\vec{k_1} + \vec{k},l} \nonumber \\
 &+ \vec{f}^{*}_{\vec{k_1} - \vec{k},l} (\vec{k},\vec{r}) e^{i E_{\vec{k_1} - \vec{k},l} t/\hbar} \hat{b}^{\dagger}_{\vec{k_1} - \vec{k},l} \nonumber \\
 &+ \vec{g}_{\vec{k_1} - \vec{k},l} (\vec{k},\vec{r}) e^{-i E_{\vec{k_1} - \vec{k},l} t/\hbar} \hat{b}_{\vec{k_1} - \vec{k},l} \nonumber \\
 &+ \vec{g}^{*}_{\vec{k_1} + \vec{k},l} (\vec{k},\vec{r}) e^{i E_{\vec{k_1} + \vec{k},l} t/\hbar} \hat{b}^{\dagger}_{\vec{k_1} + \vec{k},l} \ ,
 \label{qfluc_stripe}
\end{align}
with $\vec{k}_1 = k_1 \vec{e}_{x}$, $\vec{e}_{x}$ being the unitary vector along the $x$-axis. Notice also that, in this expression, the momentum runs over all possible values corresponding to the first Brillouin Zone ($0<k_x<2k_1$, $0<k_y,k_z<\infty$). The excitation spectrum of the system and the Bogoliubov amplitudes can be obtained by substituting Eqs.~(\ref{tdpsi_soc}) and~(\ref{qfluc_stripe}) into Eq.~(\ref{tgpe}) and solving a diagonalization problem. This can be done expanding the amplitudes of Eq.~(\ref{qfluc_stripe}) in Bloch waves~\cite{li13}. Once with the Bogoliubov amplitudes, one can numerically calculate the LHY energy correction. Due to the condensate state featuring periodic density modulations on the $x$-axis, the LHY integral implies a sum running over all Brillouin Zones. In practice though, we truncate the sum and keep only a finite number of terms. Furthermore, additional approximations must be done in order to keep the computational cost of the calculation down to a reasonable level, since the size of the aforementioned diagonalization problem scales with the number of Brillouin Zones in the integration~\autoref{sec:supplementary}.
 
As happens in the non-SOC system, the LHY integral is ultraviolet divergent, and must be regularized. By computing the LHY integral over increasingly larger cylindrical domains, we find that its divergent behavior can be fitted to that of the integral $I_{\eta}(V_I) = \int_{V_I} \vec{dk} \text{ } \eta / k^2$, with $V_I$ the integration volume and $\eta$ a fitting parameter. We use Dimensional Regularization (DR)~\cite{salasnich16,leibbrandt75} to regularize this integral~\autoref{sec:supplementary}. 
 
We are particularly interested in the role played by quantum fluctuations in the stripe phase when the mean-field system is unstable. Using the expression for the mean-field energy per particle of Ref.~\cite{li15}, and carrying on $\pdv[2]{E}{V}$, it can be shown that the 
mean-field stripe state is unstable for $G_1 < 0$ if $\abs{2 k_0^2} > \abs{G_1}$, a requirement fulfilled in all our calculations. Under these conditions, the LHY energy is positive and stabilizes the collapsing mean-field state. This is similar to the result obtained for unstable Bose-Bose mixtures without SOC~\cite{petrov15}, although remarkable differences exist between the two cases, as detailed below.
 
\begin{figure}[t]
\centering
\includegraphics[width=0.92\linewidth]{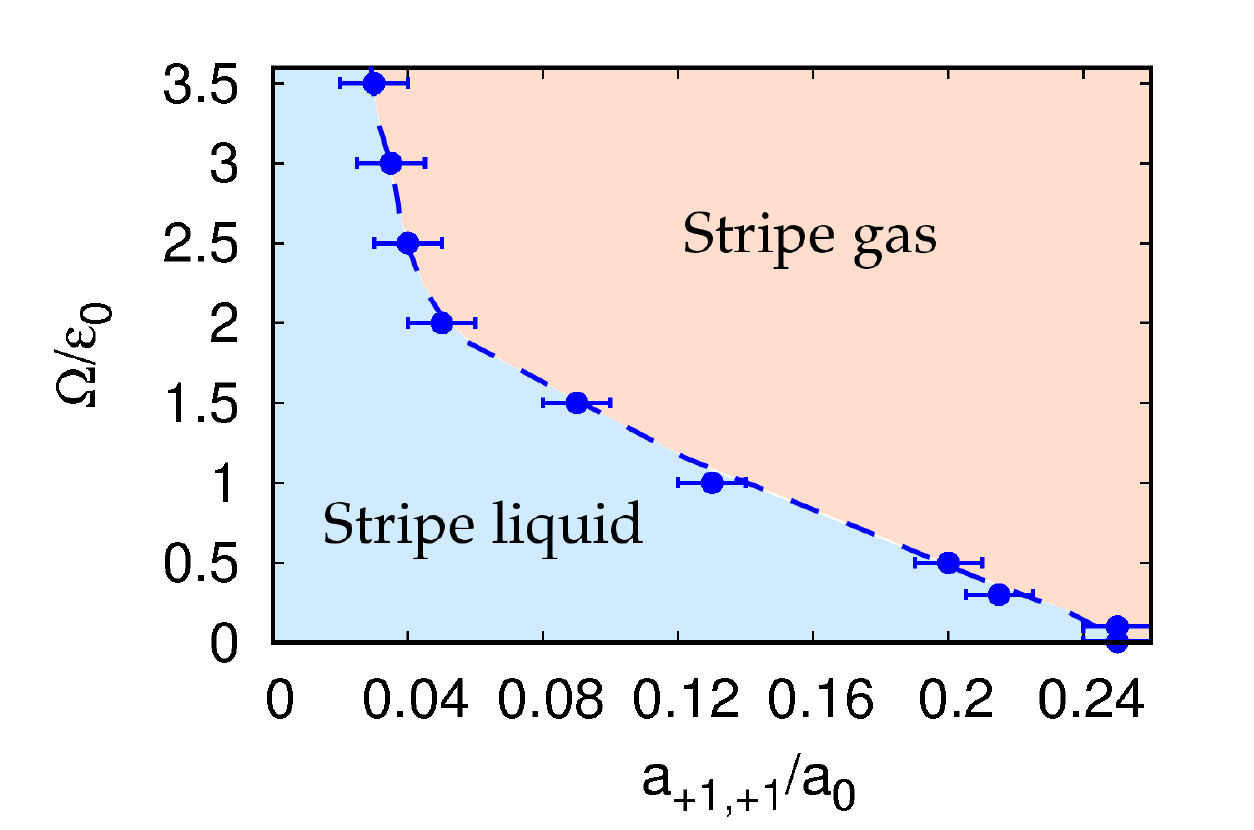}
\caption{ Phase diagram of the system stabilized by quantum fluctuations at the LHY level for the different stripe phases with $\gamma=-21$.} 
\label{fig_diagram}
\end{figure}

The phase diagram of the stabilized stripe state, as a function of $\Omega$ and $a_{+1,+1}>0$, for $\gamma=-21$ (i.e. $a_{+1,-1} = -1.1 a_{+1,+1}$) is shown in Fig.~\ref{fig_diagram}. Error bars account for the numerical error associated to the finite number of Brillouin Zones and integration points considered in the calculations. As it can be seen from the Figure, depending on the value of the Raman coupling $\Omega$ and the scattering lengths, the homogeneous system can be either a liquid ($n^{(0)} \neq 0$ with $n^{(0)} $ the density for which $E/N$ is minimum) or a gas (for which $\dv{E/N}{n} > 0$ $\forall \text{ } n$). This is an effect entirely induced by the presence of the SOC interaction, since for unstable Bose-Bose mixtures without SOC the stabilization of the collapse by the LHY energy always brings the homogeneous system to a liquid state~\cite{petrov15}. In order to determine if the system is in a liquid or in a gas state, we compute $E/N$ for different $\{ \Omega, a_{+1,+1} \}$ and densities. Typically, $n \in [3.78 \times 10^{-4} ,4.93 \times 10^{-3}]$, although this range is extended in some cases up to $n \simeq 0.1$.
 
Fig.~\ref{fig_diagram} indicates that increasing the Raman coupling leads to a lower interval of scattering lengths where the system is in the liquid phase. As a consequence, for fixed $a_{+1,+1}$, increasing $\Omega$ leads to a decrease in $n^{(0)}$, leading to a less correlated liquid. In much the same way, increasing $a_{+1,+1}$ with $\gamma=-21$ and keeping $\Omega$ constant drives the system from a liquid state to a gas, i.e., the equilibrium density $n^{(0)}$ shifts to lower values until $n^{(0)}=0$, with the \textit{gas parameter}, $n^{(0)} a_{+1,+1}^3$, also decreasing. Remarkably, this behavior is not seen in ultradilute non-SOC Bose-Bose mixtures, where  multiplying all the scattering lengths by a constant leaves $n^{(0)} a_{+1,+1}^3$ invariant~\cite{petrov15}.
 
The LHY correction reveals that the stripe phase is energetically favored with respect to the plane wave phase beyond the mean field level, within the domain  $\{ \Omega, a_{+1,+1}, n \}$ explored . Extrapolation to $\Omega$ values higher than those shown in Fig.~\ref{fig_diagram} reveals that other phases can be energetically favorable: the plane wave phase for $\Omega \gtrsim 3.8$, $a_{+1,+1} \gtrsim 0.1$, $n > 3.78 \times 10^{-4}$ (although the single minimum phase may be energetically favored for high enough densities), and the single minimum phase for $\Omega \gtrsim 4$, $a_{+1,+1} \gtrsim 0.03$, $n > 3.78 \times 10^{-4}$.

The phase diagram of Fig.~\ref{fig_diagram} has been computed fixing $a_{+1,-1} = -1.1 a_{+1,+1}$. However, due to the mean field instability present for $a_{+1,-1} < - a_{+1,+1}$, the LHY correction yields clearly unphysical imaginary contributions to the energy, since the excitation spectrum becomes imaginary at low momenta. In order to avoid that, and as usually done in the non-SOC case, we evaluate the LHY correction for $a_{+1,-1} = - a_{+1,+1}$, i.e., in the limit of the mean-field stability, while the mean-field energy terms are computed for $a_{+1,-1} < - a_{+1,+1}$. The changes in the phase diagram reported in Fig.~\ref{fig_diagram} when $E_{\text{LHY}}(a_{+1,-1} = -a_{+1,+1})$ is used instead of $\text{Re}\{ E_{\text{LHY}}(a_{+1,-1} = -1.1 a_{+1,+1}) \}$ are accounted for in the error bars.

\begin{figure}[t]
\centering
\includegraphics[width=0.85\linewidth]{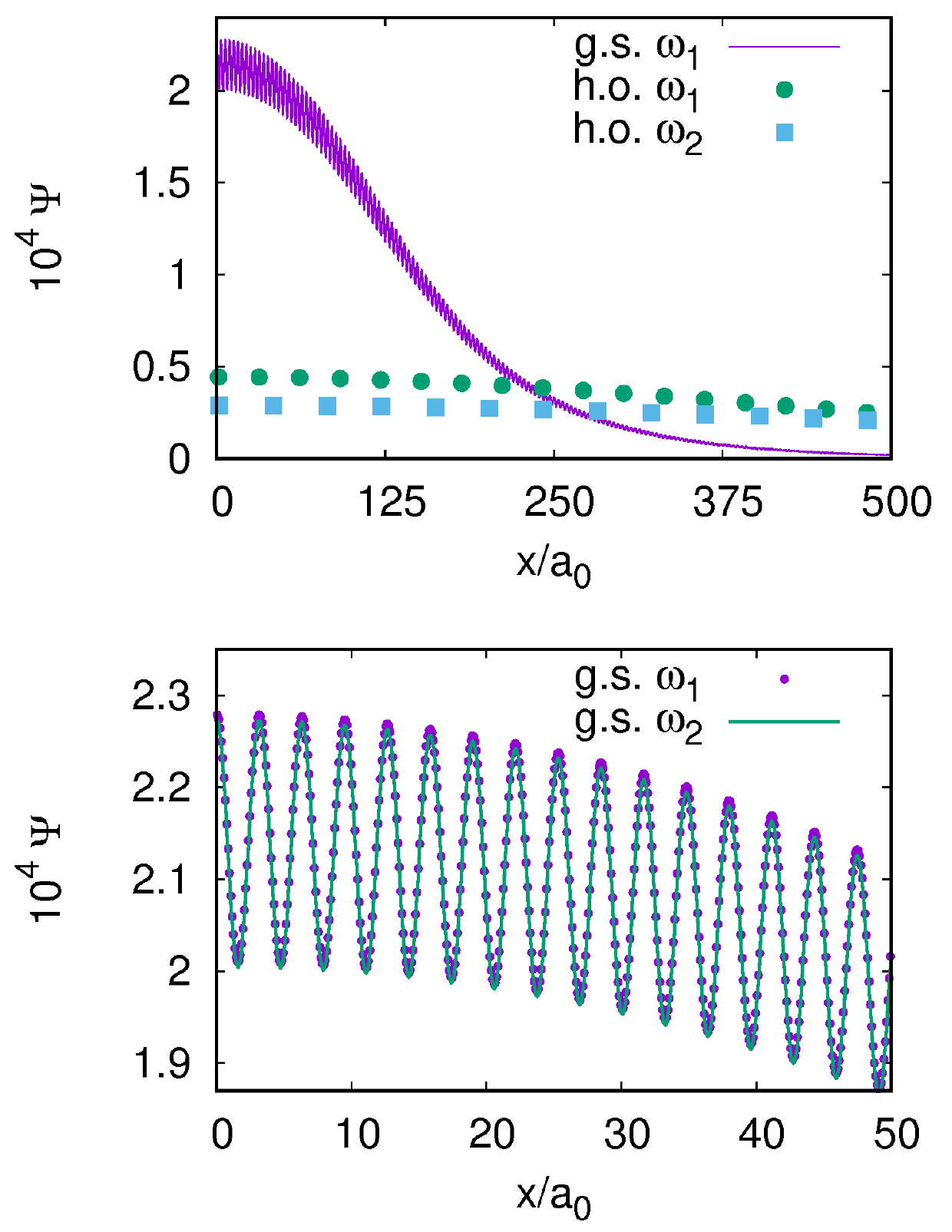}
\caption{Upper panel: density profile of the droplet along the $x$-axis for $\Omega=0.5$, $a_{+1,+1} = 0.12$, $\gamma = -21$, $N=1.4 \times 10^5$ with $\omega_1 = 4.93 \times 10^{-6}$ (blue line). The profile corresponding to the same $N$, $a_{+1,+1}$, $\gamma$, $\Omega$ values but for $\omega_2 = 2.77 \times 10^{-6}$ is indistinguishable from the one reported in the upper panel. The harmonic oscillator ground-state solution for both values of $\omega$ is shown as squares and circles. Lower panel: magnified view of the two density profiles at small $x$. Only $x>0$ values are displayed since the profile is symmetric in the $x$-axis.} 
\label{fig_droplet}
\end{figure}  
 
As it happens in ultradilute non-SOC Bose-Bose mixtures, a finite size system in the liquid stripe phase can form a droplet. However, in the SOC case, the droplets show a striped pattern along the $x$ direction, defined by $\hat{P}_x$ in $\hat{W}^{\text{SOC}}$ (\ref{Wraman}). Since Raman SOC stripes are known to be supersolid~\cite{li13,chen18,sanchez20}, the resulting striped droplets represent a novel quantum state of matter that mixes the self-bound character of liquids, the spatial periodicity present in solids and a superfluid behavior. To obtain the ground state of the finite system, we solve the extended Gross-Pitaevskii equation (eGPE). To this end,  we build a density-dependent energy functional by fitting the obtained LHY energy correction for different densities $n$. The chosen functional form is $\left[ E_{\text{LHY}}/N \right] (n) = b n + a n^{3/2}$, with $a$ and $b$ two fitting parameters, which consistently reproduces our data in the range of densities spanned in this work. In order to obtain the eGPE, we minimize $E(\Psi, \Psi^{\dagger}) = \int \vec{dr} \left( \epsilon_{\text{MF}}(\Psi, \Psi^{\dagger}) + V_{\text{osc}}(\vec{r})\abs{\Psi}^2 + \epsilon_{\text{LHY}} (\Psi, \Psi^{\dagger}) \right)$,  replacing $n \rightarrow \Psi^{\dagger} \Psi$ in the $\epsilon_{\text{LHY}}$ term. Here, $\epsilon_{\text{MF}}$ and $\epsilon_{\text{LHY}}$ are the mean-field and Lee-Huang-Yang energy densities of the infinite system, respectively, and $\Psi$ is the spinor wave function. The harmonic oscillator potential, $V_{\text{osc}}(\vec{r})= \omega^2 r^2$ in reduced units, is added to keep the system finite.

Solving directly the eGPE is technically involved for some values of the system size because of the presence of two very different length scales: on one hand, the period of the stripes, which for values of $\Omega \leq 1$ is of order $L_s \sim \order{1}$ and, on the other, the radius of the droplet, which is generally much larger. Nevertheless, results for a set of parameters, for which the problem is well conditioned, show that the ground-state wave function of the system obtained from the eGPE can be well approximated by
\begin{equation}
 \Psi(\vec{r}) \simeq f_{\text{stripe}}(x) f_{\text{droplet}} (r) \ ,
 \label{droplet_ansatz}
\end{equation}
with errors on the momentum of the stripes of at most $5 \%$. Here, $f_{\text{stripe}}(x) \simeq \psi_0(\vec{r})$, the mean-field ansatz of Ref.~\cite{li15}, which equals Eq.~\ref{stripe_cond} considering only $n=-1$ and $n=0$ in the sum. The function $f_{\text{droplet}} (r)$ depends only on $r = \abs{\vec{r}}$, with $\vec{r}$ the position vector in three dimensions. In order to efficiently calculate $f_{\text{droplet}} (r)$, we apply a further approximation: we solve the eGPE obtained from the functional $\tilde{E}(\Psi, \Psi^{\dagger}) = \int \vec{dr} \left( \tilde{\epsilon}_{\text{MF}}(\Psi, \Psi^{\dagger}) + V_{\text{osc}}(\vec{r})\abs{\Psi}^2 + \epsilon_{\text{LHY, SOC}} (\Psi, \Psi^{\dagger}) \right)$. Here, $\tilde{\epsilon}_{\text{MF}}$ is the mean-field energy density obtained with the SOC terms removed, while $\epsilon_{\text{LHY, SOC}}$ is the LHY energy density obtained from the full SOC calculation. Then, the resulting eGPE can be solved efficiently as the problem only depends on $r$.

There is a minimum particle number required for  having a stable self-bound droplet in the ground state  which is known as  critical number, $N_{\text{crit}}$. We determine $N_{\text{crit}}$ by solving the eGPE for different strengths of the trapping potential and comparing the solution obtained to the ground-state wave function of the harmonic oscillator. For $N \geq N_{\text{crit}}$, changing the trapping strength leaves the solution of the eGPE unaffected. Also, the energy of the droplet state must fulfill the condition $\tilde{E} < 0$ for low enough $\omega$. We show in Fig.~\ref{fig_droplet} the function $\tilde{\Psi}(\vec{r}) = \left( f_{\text{stripe}}(x) f_{\text{droplet}} (r) \right)/\sqrt{ \int \vec{dr} f^2_{\text{droplet}}(r) }$ along the $x$-axis corresponding to a case where a stable droplet is formed, with parameters $\Omega=0.5$, $a_{+1,+1} = 0.12$, $\gamma = -21$, $N=1.4 \times 10^5$. The trapping strengths are $\omega_1 =4.93 \times 10^{-6}$ and $\omega_2 = 2.77 \times 10^{-6}$. We only show the $+1$ spinor component, since $\Psi^{+1}(\vec{r}) = \Psi^{-1}(\vec{r})$. As one can see in Fig.~\ref{fig_droplet}, the density profile of the droplet clearly manifests oscillations in the density, characteristic of the stripe phase.

\begin{figure}[b]
\centering
\includegraphics[width=0.92\linewidth]{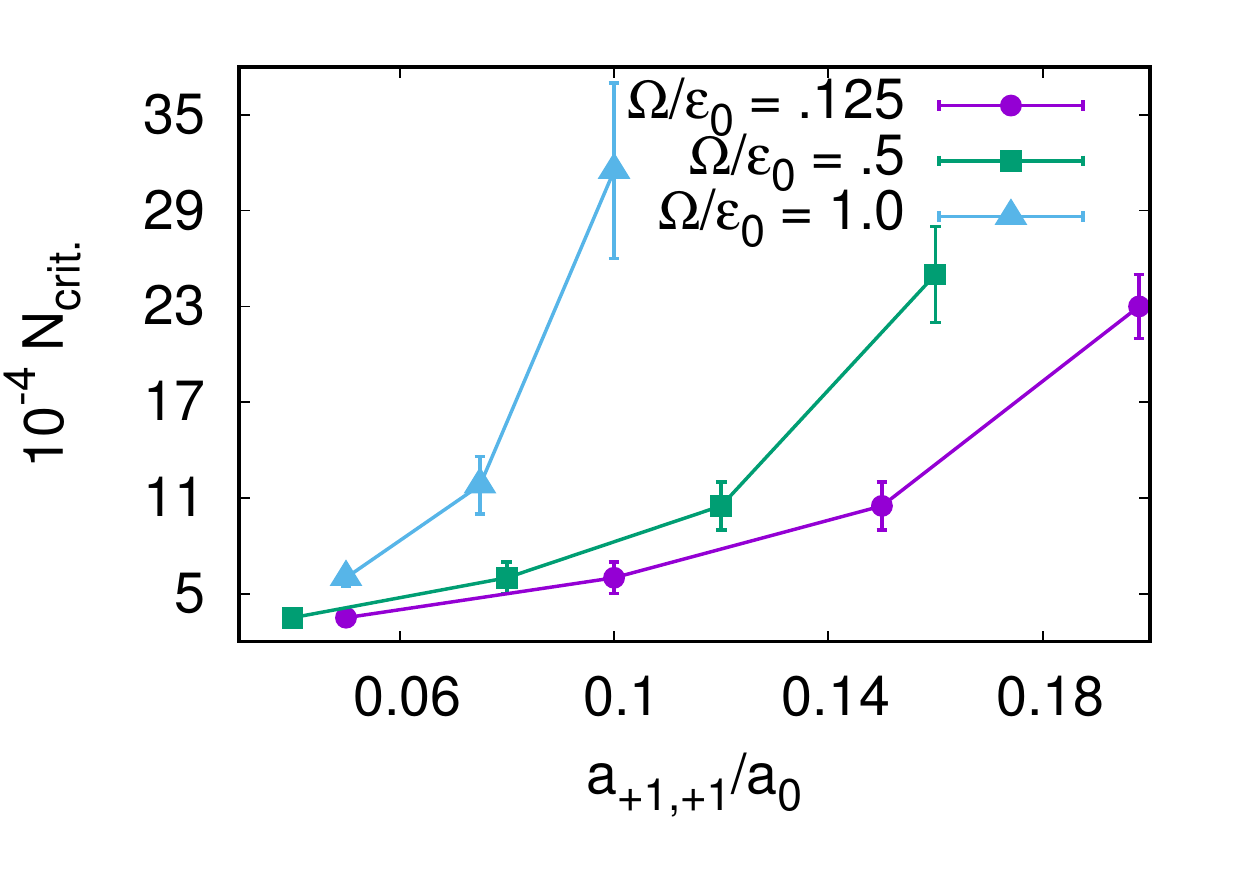}
\caption{Critical number as a function of the scattering length $a_{+1,+1}$ for $\gamma=-21$ ($a_{+1,-1} = -1.1 a_{+1,+1}$), for different values of $\Omega$. Lines are a guide to the eye.} 
\label{fig_crit_number}
\end{figure}  

We report in Fig.~\ref{fig_crit_number}, the critical number as a function of $a_{+1,+1}$ for $\Omega = 0.125\text{, }0.5\text{, }1.0$ and $\gamma=-21$ ($a_{+1,-1} = -1.1 a_{+1,+1}$). Errorbars account for the numerical inaccuracies associated to both the finite number of Brillouin Zones being integrated and the number of points used in the computation of $E_{\text{LHY}}$, and also for the difference in the results obtained when employing either $E_{\text{LHY}}(a_{+1,-1} = -a_{+1,+1})$ or $E_{\text{LHY}}(a_{+1,-1} = -1.1 a_{+1,+1})$. As can be seen from the Figure, the critical number increases with both $\Omega$ and the scattering lengths, consistently with the results shown in Fig.~\ref{fig_diagram}. Remarkably, the critical numbers obtained are reachable in current experimental setups, opening the possibility to observe and measure quantum properties of striped droplets. For the sake of comparison, previous experiments with SOC systems have been carried out with $N \sim 4 \times 10^5$~\cite{spielman11} and $N \sim 1.4 \times 10^5$~\cite{ketterle17} particles. Another interesting quantity regarding the number of particles of a droplet is the saturation number, $N_s$. If $N > N_s$, $f_{\text{droplet}}(r)$ shows a plateau at a range of positions $r \in [0\text{, }r_{\text{max.}}]$ with $r_{\text{max.}}$ increasing as $N$ increases. For $\Omega = 0.125$, the saturation number in all cases is of $\order{10^6}$ or higher, which makes it challenging to be observed.

Despite the evaluation of $E_{\text{LHY}}$ for SOC systems presented in this work is quite more elaborate than in non-SOC systems, the resulting observed dependence on the system parameters is smooth enough to allow for a simple functional form approximation. In this way, we report an approximated density functional for $E_{\text{LHY}}(a_{+1,-1} = -a_{+1,+1})$ in the stripe phase. This functional depends on $a_{+1,+1}$, $n$ and $\Omega$, and has been obtained by fitting the LHY energies in different density regimes. It is given by
\begin{equation}
 \eval{E_{\text{LHY}}/N}_{ \substack{ a_{+1,-1} =\\ -a_{+1,+1} } } \simeq (A + B\Omega^2) n a_{+1,+1}^2 + C \sqrt{ n^3 a_{+1,+1}^5 } \ ,
 \label{e_lhy_functional}
\end{equation}
with $A=1.89 \pm 0.04$, $B=2.17 \pm 0.03$ and $C=37 \pm 2$ in dimensionless form. The above expression reproduces the obtained LHY energies with errors between $1 \%$ and $10 \%$ for $0 < \Omega < 3$, $0 < n \lesssim 0.1$, $0 < a_{+1,+1} < 0.225$, although the limiting value of the density can be increased further for scattering lengths $a_{+1,+1} \lesssim 0.05$, keeping the error of the functional approximation within the mentioned boundaries. 

The are several reasons behind the choice of the functional form in Eq.~\ref{e_lhy_functional}. The functional features a difference of one between the exponents of the scattering length and the density in each term, which is required. Also, the linear dependence with respect to the density can be clearly observed at low density regimes. The term proportional to $\sqrt{ n^3 a_{+1,+1}^5 }$ has been chosen in analogy with the non-SOC case. Finally, the fitting process reveals that higher order terms with respect to the density and the scattering length are irrelevant in the density regimes considered in this work. 
 
In conclusion, we have evaluated the role of quantum fluctuations in a striped system under Raman SOC that is unstable at the mean-field level. We have found that quantum fluctuations prevent the mean field collapse as happens in regular ultradilute non-SOC Bose-Bose mixtures. However, the presence of SOC induces the emergence of two stable phases: a gas phase and a liquid phase, with the Raman coupling and the scattering lengths determining the one that is energetically favorable. The liquid phase of this system represents a state of matter which shows superfluidity and periodicity along one direction. We have evaluated the ground state of the finite system by solving the eGPE to find self-bound droplet-like solutions with periodicity along the $x$-axis as a result. These droplets represent a novel state of matter that combines the self-bound character of liquids, a density modulation and superfluidity. We have also computed the critical numbers associated to the self-bound droplet states and found that they are experimentally accessible. Finally, we have provided an approximated energy functional for the Lee-Huang-Yang energy in the stripe phase. We hope that this work can inspire new experiments to detect the proposed novel supersolid striped droplets. 
 
We acknowledge L. Tarruell and V. Cikojević for fruitful discussions. This work has been supported by the MINECO (Spain) Grant No. 
FIS2017-84114-C2-1-P. J. S\'anchez-Baena also acknowledges the FPU fellowship with reference FPU15/01805 from MCIU.

\bibliography{SOC_LHY}

\begin{thebibliography}{21}%
\makeatletter
\providecommand \@ifxundefined [1]{%
 \@ifx{#1\undefined}
}%
\providecommand \@ifnum [1]{%
 \ifnum #1\expandafter \@firstoftwo
 \else \expandafter \@secondoftwo
 \fi
}%
\providecommand \@ifx [1]{%
 \ifx #1\expandafter \@firstoftwo
 \else \expandafter \@secondoftwo
 \fi
}%
\providecommand \natexlab [1]{#1}%
\providecommand \enquote  [1]{``#1''}%
\providecommand \bibnamefont  [1]{#1}%
\providecommand \bibfnamefont [1]{#1}%
\providecommand \citenamefont [1]{#1}%
\providecommand \href@noop [0]{\@secondoftwo}%
\providecommand \href [0]{\begingroup \@sanitize@url \@href}%
\providecommand \@href[1]{\@@startlink{#1}\@@href}%
\providecommand \@@href[1]{\endgroup#1\@@endlink}%
\providecommand \@sanitize@url [0]{\catcode `\\12\catcode `\$12\catcode
  `\&12\catcode `\#12\catcode `\^12\catcode `\_12\catcode `\%12\relax}%
\providecommand \@@startlink[1]{}%
\providecommand \@@endlink[0]{}%
\providecommand \url  [0]{\begingroup\@sanitize@url \@url }%
\providecommand \@url [1]{\endgroup\@href {#1}{\urlprefix }}%
\providecommand \urlprefix  [0]{URL }%
\providecommand \Eprint [0]{\href }%
\providecommand \doibase [0]{http://dx.doi.org/}%
\providecommand \selectlanguage [0]{\@gobble}%
\providecommand \bibinfo  [0]{\@secondoftwo}%
\providecommand \bibfield  [0]{\@secondoftwo}%
\providecommand \translation [1]{[#1]}%
\providecommand \BibitemOpen [0]{}%
\providecommand \bibitemStop [0]{}%
\providecommand \bibitemNoStop [0]{.\EOS\space}%
\providecommand \EOS [0]{\spacefactor3000\relax}%
\providecommand \BibitemShut  [1]{\csname bibitem#1\endcsname}%
\let\auto@bib@innerbib\@empty
\bibitem [{\citenamefont {Hasan}\ and\ \citenamefont {Kane}(2010)}]{hasan10}%
  \BibitemOpen
  \bibfield  {author} {\bibinfo {author} {\bibfnamefont {M.~Z.}\ \bibnamefont
  {Hasan}}\ and\ \bibinfo {author} {\bibfnamefont {C.~L.}\ \bibnamefont
  {Kane}},\ }\href {\doibase 10.1103/RevModPhys.82.3045} {\bibfield  {journal}
  {\bibinfo  {journal} {Rev. Mod. Phys.}\ }\textbf {\bibinfo {volume} {82}},\
  \bibinfo {pages} {3045--3067} (\bibinfo {year} {2010})}\BibitemShut {NoStop}%
\bibitem [{\citenamefont {Sato}\ and\ \citenamefont {Ando}(2017)}]{sato17}%
  \BibitemOpen
  \bibfield  {author} {\bibinfo {author} {\bibfnamefont {Masatoshi}\
  \bibnamefont {Sato}}\ and\ \bibinfo {author} {\bibfnamefont {Yoichi}\
  \bibnamefont {Ando}},\ }\href {\doibase 10.1088/1361-6633/aa6ac7} {\bibfield
  {journal} {\bibinfo  {journal} {Reports on Progress in Physics}\ }\textbf
  {\bibinfo {volume} {80}},\ \bibinfo {pages} {076501} (\bibinfo {year}
  {2017})}\BibitemShut {NoStop}%
\bibitem [{\citenamefont {Zhang}\ and\ \citenamefont {Liu}(2018)}]{zhang18}%
  \BibitemOpen
  \bibfield  {author} {\bibinfo {author} {\bibfnamefont {Long}\ \bibnamefont
  {Zhang}}\ and\ \bibinfo {author} {\bibfnamefont {Xiong-Jun}\ \bibnamefont
  {Liu}},\ }\enquote {\bibinfo {title} {Review article: Spin-orbit coupling and
  topological phases for ultracold atoms},}\ \ (\bibinfo {year} {2018})\ pp.\
  \bibinfo {pages} {1--87}\BibitemShut {NoStop}%
\bibitem [{\citenamefont {Lin}\ \emph {et~al.}(2011)\citenamefont {Lin},
  \citenamefont {Jim{\'e}nez-Garc{\'i}a},\ and\ \citenamefont
  {Spielman}}]{spielman11}%
  \BibitemOpen
  \bibfield  {author} {\bibinfo {author} {\bibfnamefont {Y.-J.}\ \bibnamefont
  {Lin}}, \bibinfo {author} {\bibfnamefont {K.}~\bibnamefont
  {Jim{\'e}nez-Garc{\'i}a}}, \ and\ \bibinfo {author} {\bibfnamefont {I.~B.}\
  \bibnamefont {Spielman}},\ }\href {\doibase 10.1038/nature09887} {\bibfield
  {journal} {\bibinfo  {journal} {Nature}\ }\textbf {\bibinfo {volume} {471}},\
  \bibinfo {pages} {83--86} (\bibinfo {year} {2011})}\BibitemShut {NoStop}%
\bibitem [{\citenamefont {Li}\ \emph {et~al.}()\citenamefont {Li},
  \citenamefont {Martone},\ and\ \citenamefont {Stringari}}]{li15}%
  \BibitemOpen
  \bibfield  {author} {\bibinfo {author} {\bibfnamefont {Yun}\ \bibnamefont
  {Li}}, \bibinfo {author} {\bibfnamefont {Giovanni~I.}\ \bibnamefont
  {Martone}}, \ and\ \bibinfo {author} {\bibfnamefont {Sandro}\ \bibnamefont
  {Stringari}},\ }in\ \href {\doibase 10.1142/9789814667746_0005} {\emph
  {\bibinfo {booktitle} {Annual Review of Cold Atoms and Molecules}}},\ Chap.\
  \bibinfo {chapter} {CHAPTER 5}, pp.\ \bibinfo {pages} {201--250}\BibitemShut
  {NoStop}%
\bibitem [{\citenamefont {Hamner}\ \emph {et~al.}(2015)\citenamefont {Hamner},
  \citenamefont {Zhang}, \citenamefont {Khamehchi}, \citenamefont {Davis},\
  and\ \citenamefont {Engels}}]{hamner15}%
  \BibitemOpen
  \bibfield  {author} {\bibinfo {author} {\bibfnamefont {C.}~\bibnamefont
  {Hamner}}, \bibinfo {author} {\bibfnamefont {Yongping}\ \bibnamefont
  {Zhang}}, \bibinfo {author} {\bibfnamefont {M.~A.}\ \bibnamefont
  {Khamehchi}}, \bibinfo {author} {\bibfnamefont {Matthew~J.}\ \bibnamefont
  {Davis}}, \ and\ \bibinfo {author} {\bibfnamefont {P.}~\bibnamefont
  {Engels}},\ }\href {\doibase 10.1103/PhysRevLett.114.070401} {\bibfield
  {journal} {\bibinfo  {journal} {Phys. Rev. Lett.}\ }\textbf {\bibinfo
  {volume} {114}},\ \bibinfo {pages} {070401} (\bibinfo {year}
  {2015})}\BibitemShut {NoStop}%
\bibitem [{\citenamefont {Bersano}\ \emph {et~al.}(2019)\citenamefont
  {Bersano}, \citenamefont {Hou}, \citenamefont {Mossman}, \citenamefont
  {Gokhroo}, \citenamefont {Luo}, \citenamefont {Sun}, \citenamefont {Zhang},\
  and\ \citenamefont {Engels}}]{bersano19}%
  \BibitemOpen
  \bibfield  {author} {\bibinfo {author} {\bibfnamefont {Thomas~M.}\
  \bibnamefont {Bersano}}, \bibinfo {author} {\bibfnamefont {Junpeng}\
  \bibnamefont {Hou}}, \bibinfo {author} {\bibfnamefont {Sean}\ \bibnamefont
  {Mossman}}, \bibinfo {author} {\bibfnamefont {Vandna}\ \bibnamefont
  {Gokhroo}}, \bibinfo {author} {\bibfnamefont {Xi-Wang}\ \bibnamefont {Luo}},
  \bibinfo {author} {\bibfnamefont {Kuei}\ \bibnamefont {Sun}}, \bibinfo
  {author} {\bibfnamefont {Chuanwei}\ \bibnamefont {Zhang}}, \ and\ \bibinfo
  {author} {\bibfnamefont {Peter}\ \bibnamefont {Engels}},\ }\href {\doibase
  10.1103/PhysRevA.99.051602} {\bibfield  {journal} {\bibinfo  {journal} {Phys.
  Rev. A}\ }\textbf {\bibinfo {volume} {99}},\ \bibinfo {pages} {051602}
  (\bibinfo {year} {2019})}\BibitemShut {NoStop}%
\bibitem [{\citenamefont {Cheuk}\ \emph {et~al.}(2012)\citenamefont {Cheuk},
  \citenamefont {Sommer}, \citenamefont {Hadzibabic}, \citenamefont {Yefsah},
  \citenamefont {Bakr},\ and\ \citenamefont {Zwierlein}}]{cheuk12}%
  \BibitemOpen
  \bibfield  {author} {\bibinfo {author} {\bibfnamefont {Lawrence~W.}\
  \bibnamefont {Cheuk}}, \bibinfo {author} {\bibfnamefont {Ariel~T.}\
  \bibnamefont {Sommer}}, \bibinfo {author} {\bibfnamefont {Zoran}\
  \bibnamefont {Hadzibabic}}, \bibinfo {author} {\bibfnamefont {Tarik}\
  \bibnamefont {Yefsah}}, \bibinfo {author} {\bibfnamefont {Waseem~S.}\
  \bibnamefont {Bakr}}, \ and\ \bibinfo {author} {\bibfnamefont {Martin~W.}\
  \bibnamefont {Zwierlein}},\ }\href {\doibase 10.1103/PhysRevLett.109.095302}
  {\bibfield  {journal} {\bibinfo  {journal} {Phys. Rev. Lett.}\ }\textbf
  {\bibinfo {volume} {109}},\ \bibinfo {pages} {095302} (\bibinfo {year}
  {2012})}\BibitemShut {NoStop}%
\bibitem [{\citenamefont {Wang}\ \emph {et~al.}(2012)\citenamefont {Wang},
  \citenamefont {Yu}, \citenamefont {Fu}, \citenamefont {Miao}, \citenamefont
  {Huang}, \citenamefont {Chai}, \citenamefont {Zhai},\ and\ \citenamefont
  {Zhang}}]{wang12}%
  \BibitemOpen
  \bibfield  {author} {\bibinfo {author} {\bibfnamefont {Pengjun}\ \bibnamefont
  {Wang}}, \bibinfo {author} {\bibfnamefont {Zeng-Qiang}\ \bibnamefont {Yu}},
  \bibinfo {author} {\bibfnamefont {Zhengkun}\ \bibnamefont {Fu}}, \bibinfo
  {author} {\bibfnamefont {Jiao}\ \bibnamefont {Miao}}, \bibinfo {author}
  {\bibfnamefont {Lianghui}\ \bibnamefont {Huang}}, \bibinfo {author}
  {\bibfnamefont {Shijie}\ \bibnamefont {Chai}}, \bibinfo {author}
  {\bibfnamefont {Hui}\ \bibnamefont {Zhai}}, \ and\ \bibinfo {author}
  {\bibfnamefont {Jing}\ \bibnamefont {Zhang}},\ }\href {\doibase
  10.1103/PhysRevLett.109.095301} {\bibfield  {journal} {\bibinfo  {journal}
  {Phys. Rev. Lett.}\ }\textbf {\bibinfo {volume} {109}},\ \bibinfo {pages}
  {095301} (\bibinfo {year} {2012})}\BibitemShut {NoStop}%
\bibitem [{\citenamefont {Li}\ \emph {et~al.}(2013)\citenamefont {Li},
  \citenamefont {Martone}, \citenamefont {Pitaevskii},\ and\ \citenamefont
  {Stringari}}]{li13}%
  \BibitemOpen
  \bibfield  {author} {\bibinfo {author} {\bibfnamefont {Yun}\ \bibnamefont
  {Li}}, \bibinfo {author} {\bibfnamefont {Giovanni~I.}\ \bibnamefont
  {Martone}}, \bibinfo {author} {\bibfnamefont {Lev~P.}\ \bibnamefont
  {Pitaevskii}}, \ and\ \bibinfo {author} {\bibfnamefont {Sandro}\ \bibnamefont
  {Stringari}},\ }\href {\doibase 10.1103/PhysRevLett.110.235302} {\bibfield
  {journal} {\bibinfo  {journal} {Phys. Rev. Lett.}\ }\textbf {\bibinfo
  {volume} {110}},\ \bibinfo {pages} {235302} (\bibinfo {year}
  {2013})}\BibitemShut {NoStop}%
\bibitem [{\citenamefont {Macia}\ \emph {et~al.}(2014)\citenamefont {Macia},
  \citenamefont {Boronat},\ and\ \citenamefont {Mazzanti}}]{macia14}%
  \BibitemOpen
  \bibfield  {author} {\bibinfo {author} {\bibfnamefont {A.}~\bibnamefont
  {Macia}}, \bibinfo {author} {\bibfnamefont {J.}~\bibnamefont {Boronat}}, \
  and\ \bibinfo {author} {\bibfnamefont {F.}~\bibnamefont {Mazzanti}},\ }\href
  {\doibase 10.1103/PhysRevA.90.061601} {\bibfield  {journal} {\bibinfo
  {journal} {Phys. Rev. A}\ }\textbf {\bibinfo {volume} {90}},\ \bibinfo
  {pages} {061601} (\bibinfo {year} {2014})}\BibitemShut {NoStop}%
\bibitem [{\citenamefont {Bombin}\ \emph {et~al.}(2017)\citenamefont {Bombin},
  \citenamefont {Boronat},\ and\ \citenamefont {Mazzanti}}]{bombin17}%
  \BibitemOpen
  \bibfield  {author} {\bibinfo {author} {\bibfnamefont {R.}~\bibnamefont
  {Bombin}}, \bibinfo {author} {\bibfnamefont {J.}~\bibnamefont {Boronat}}, \
  and\ \bibinfo {author} {\bibfnamefont {F.}~\bibnamefont {Mazzanti}},\ }\href
  {\doibase 10.1103/PhysRevLett.119.250402} {\bibfield  {journal} {\bibinfo
  {journal} {Phys. Rev. Lett.}\ }\textbf {\bibinfo {volume} {119}},\ \bibinfo
  {pages} {250402} (\bibinfo {year} {2017})}\BibitemShut {NoStop}%
\bibitem [{\citenamefont {Li}\ \emph {et~al.}(2017)\citenamefont {Li},
  \citenamefont {Lee}, \citenamefont {Huang}, \citenamefont {Burchesky},
  \citenamefont {Shteynas}, \citenamefont {Top}, \citenamefont {Jamison},\ and\
  \citenamefont {Ketterle}}]{ketterle17}%
  \BibitemOpen
  \bibfield  {author} {\bibinfo {author} {\bibfnamefont {Jun-Ru}\ \bibnamefont
  {Li}}, \bibinfo {author} {\bibfnamefont {Jeongwon}\ \bibnamefont {Lee}},
  \bibinfo {author} {\bibfnamefont {Wujie}\ \bibnamefont {Huang}}, \bibinfo
  {author} {\bibfnamefont {Sean}\ \bibnamefont {Burchesky}}, \bibinfo {author}
  {\bibfnamefont {Boris}\ \bibnamefont {Shteynas}}, \bibinfo {author}
  {\bibfnamefont {Furkan~{\c{C}}a{\u{g}}r{\i}}\ \bibnamefont {Top}}, \bibinfo
  {author} {\bibfnamefont {Alan~O.}\ \bibnamefont {Jamison}}, \ and\ \bibinfo
  {author} {\bibfnamefont {Wolfgang}\ \bibnamefont {Ketterle}},\ }\href
  {\doibase 10.1038/nature21431} {\bibfield  {journal} {\bibinfo  {journal}
  {Nature}\ }\textbf {\bibinfo {volume} {543}},\ \bibinfo {pages} {91--94}
  (\bibinfo {year} {2017})}\BibitemShut {NoStop}%
\bibitem [{\citenamefont {Chen}\ \emph {et~al.}(2018)\citenamefont {Chen},
  \citenamefont {Wang}, \citenamefont {Li}, \citenamefont {Liu},\ and\
  \citenamefont {Hu}}]{chen18}%
  \BibitemOpen
  \bibfield  {author} {\bibinfo {author} {\bibfnamefont {Xiao-Long}\
  \bibnamefont {Chen}}, \bibinfo {author} {\bibfnamefont {Jia}\ \bibnamefont
  {Wang}}, \bibinfo {author} {\bibfnamefont {Yun}\ \bibnamefont {Li}}, \bibinfo
  {author} {\bibfnamefont {Xia-Ji}\ \bibnamefont {Liu}}, \ and\ \bibinfo
  {author} {\bibfnamefont {Hui}\ \bibnamefont {Hu}},\ }\href {\doibase
  10.1103/PhysRevA.98.013614} {\bibfield  {journal} {\bibinfo  {journal} {Phys.
  Rev. A}\ }\textbf {\bibinfo {volume} {98}},\ \bibinfo {pages} {013614}
  (\bibinfo {year} {2018})}\BibitemShut {NoStop}%
\bibitem [{\citenamefont {S\'anchez-Baena}\ \emph {et~al.}(2020)\citenamefont
  {S\'anchez-Baena}, \citenamefont {Boronat},\ and\ \citenamefont
  {Mazzanti}}]{sanchez20}%
  \BibitemOpen
  \bibfield  {author} {\bibinfo {author} {\bibfnamefont {J.}~\bibnamefont
  {S\'anchez-Baena}}, \bibinfo {author} {\bibfnamefont {J.}~\bibnamefont
  {Boronat}}, \ and\ \bibinfo {author} {\bibfnamefont {F.}~\bibnamefont
  {Mazzanti}},\ }\href {\doibase 10.1103/PhysRevA.101.043602} {\bibfield
  {journal} {\bibinfo  {journal} {Phys. Rev. A}\ }\textbf {\bibinfo {volume}
  {101}},\ \bibinfo {pages} {043602} (\bibinfo {year} {2020})}\BibitemShut
  {NoStop}%
\bibitem [{\citenamefont {Petrov}(2015)}]{petrov15}%
  \BibitemOpen
  \bibfield  {author} {\bibinfo {author} {\bibfnamefont {D.~S.}\ \bibnamefont
  {Petrov}},\ }\href {\doibase 10.1103/PhysRevLett.115.155302} {\bibfield
  {journal} {\bibinfo  {journal} {Phys. Rev. Lett.}\ }\textbf {\bibinfo
  {volume} {115}},\ \bibinfo {pages} {155302} (\bibinfo {year}
  {2015})}\BibitemShut {NoStop}%
\bibitem [{\citenamefont {Martone}\ \emph {et~al.}(2012)\citenamefont
  {Martone}, \citenamefont {Li}, \citenamefont {Pitaevskii},\ and\
  \citenamefont {Stringari}}]{martone12}%
  \BibitemOpen
  \bibfield  {author} {\bibinfo {author} {\bibfnamefont {Giovanni~I.}\
  \bibnamefont {Martone}}, \bibinfo {author} {\bibfnamefont {Yun}\ \bibnamefont
  {Li}}, \bibinfo {author} {\bibfnamefont {Lev~P.}\ \bibnamefont {Pitaevskii}},
  \ and\ \bibinfo {author} {\bibfnamefont {Sandro}\ \bibnamefont {Stringari}},\
  }\href {\doibase 10.1103/PhysRevA.86.063621} {\bibfield  {journal} {\bibinfo
  {journal} {Phys. Rev. A}\ }\textbf {\bibinfo {volume} {86}},\ \bibinfo
  {pages} {063621} (\bibinfo {year} {2012})}\BibitemShut {NoStop}%
\bibitem [{\citenamefont {Martone}\ and\ \citenamefont
  {Shlyapnikov}(2018)}]{martone18}%
  \BibitemOpen
  \bibfield  {author} {\bibinfo {author} {\bibfnamefont {G.~I.}\ \bibnamefont
  {Martone}}\ and\ \bibinfo {author} {\bibfnamefont {G.~V.}\ \bibnamefont
  {Shlyapnikov}},\ }\href {\doibase 10.1134/S1063776118110146} {\bibfield
  {journal} {\bibinfo  {journal} {Journal of Experimental and Theoretical
  Physics}\ }\textbf {\bibinfo {volume} {127}},\ \bibinfo {pages} {865--876}
  (\bibinfo {year} {2018})}\BibitemShut {NoStop}%
\bibitem [{\citenamefont {Harland}\ and\ \citenamefont
  {Salamon}(1988)}]{harland88}%
  \BibitemOpen
  \bibfield  {author} {\bibinfo {author} {\bibfnamefont {John~Ro}\ \bibnamefont
  {Harland}}\ and\ \bibinfo {author} {\bibfnamefont {Peter}\ \bibnamefont
  {Salamon}},\ }\href {\doibase https://doi.org/10.1016/0920-5632(88)90023-0}
  {\bibfield  {journal} {\bibinfo  {journal} {Nuclear Physics B - Proceedings
  Supplements}\ }\textbf {\bibinfo {volume} {5}},\ \bibinfo {pages} {109 --
  115} (\bibinfo {year} {1988})}\BibitemShut {NoStop}%
\bibitem [{\citenamefont {Salasnich}\ and\ \citenamefont
  {Toigo}(2016)}]{salasnich16}%
  \BibitemOpen
  \bibfield  {author} {\bibinfo {author} {\bibfnamefont {Luca}\ \bibnamefont
  {Salasnich}}\ and\ \bibinfo {author} {\bibfnamefont {Flavio}\ \bibnamefont
  {Toigo}},\ }\href {\doibase https://doi.org/10.1016/j.physrep.2016.06.003}
  {\bibfield  {journal} {\bibinfo  {journal} {Physics Reports}\ }\textbf
  {\bibinfo {volume} {640}},\ \bibinfo {pages} {1 -- 29} (\bibinfo {year}
  {2016})}\BibitemShut {NoStop}%
\bibitem [{\citenamefont {Leibbrandt}(1975)}]{leibbrandt75}%
  \BibitemOpen
  \bibfield  {author} {\bibinfo {author} {\bibfnamefont {George}\ \bibnamefont
  {Leibbrandt}},\ }\href {\doibase 10.1103/RevModPhys.47.849} {\bibfield
  {journal} {\bibinfo  {journal} {Rev. Mod. Phys.}\ }\textbf {\bibinfo {volume}
  {47}},\ \bibinfo {pages} {849--876} (\bibinfo {year} {1975})}\BibitemShut
  {NoStop}%
\end{thebibliography}%




\widetext

\newpage

\phantomsection\label{sec:supplementary}
 \centerline{\Large \textbf{SUPPLEMENTARY INFORMATION}}

\vspace{0.5cm}

\noindent
\large{\textbf{Approximations on the numerical computation of the Lee-Huang-Yang integral}}

\vspace{0.5cm}

The unregularized Lee-Huang-Yang energy per particle is given by:
\begin{align}
 &E_{\text{LHY}}^{\text{unreg}}/N = \frac{1}{n} \left( I_1 + I_2 + I_3 \right)
 \label{int_lhy_stripe}
 \\
 &I_1 = \sum_{\substack{l,l' \\ s_1,s_2}} \frac{1}{(2 \pi)^3} \int_{\substack{0<k_x<k_1 \\ 0<k_y,k_z<\infty }} \vec{dk} \left\{ f_{\vec{k_1} - \vec{k},l,l',s_1}   f^{*}_{\vec{k_1} - \vec{k},l,l', s_2}   \left[ H_0(\vec{k}_1+\vec{k}+2 l' \vec{k_1},s_1,s_2) - \delta_{s_1,s_2} \mu \right] \right. \nonumber \\
 &\left. + g_{\vec{k_1} + \vec{k},l,l',s_1}   g^{*}_{\vec{k_1} + \vec{k},l,l',s_2}   \left[ H_0(\vec{k}_1-\vec{k}+2 l' \vec{k_1},s_1,s_2) - \delta_{s_1,s_2} \mu \right]  \right\}
 \label{int_lhy_stripe_aux_0} \\
 &I_2 = \frac{4 \pi \hbar^2 n}{(2 \pi)^3 m} \sum_{\substack{l \\ n_1,n_2,n_3,n_4 \\ s_1, s_2}} \int_{\substack{0<k_x<k_1 \\ 0<k_y,k_z<\infty}} \vec{dk} \left\{ \psi_{0,n_1,s_1}^{*}  \psi_{0,n_2,s_2}^{*} a_{s_1,s_2} \right. \nonumber
 \\
 & \left. \cross \left( g^{*}_{\vec{k_1} + \vec{k},l,n_3,s_1} f_{\vec{k_1}+ \vec{k},l,n_4,s_2} + f^{*}_{\vec{k_1} - \vec{k},l,n_3,s_1} g_{\vec{k_1} - \vec{k},l,n_4,s_2}  \right) \delta[n_3+n_4-n_1-n_2] \right\}
 \label{int_lhy_stripe_aux_1}
 \\
 &I_3 = \frac{4 \pi \hbar^2 n}{(2 \pi)^3 m} \sum_{\substack{l \\ n_1,n_2,n_3,n_4 \\ s_1, s_2}} \int_{\substack{0<k_x<k_1 \\ 0<k_y,k_z<\infty}} \vec{dk} \psi_{0,n_1,s_1}^{*}  \psi_{0,n_2,s_1} a_{s_1,s_2} \left( g_{\vec{k_1} + \vec{k},l,n_3,s_2} g^{*}_{\vec{k_1} + \vec{k},l,n_4,s_2} + f^{*}_{\vec{k_1} - \vec{k},l,n_3,s_2} f_{\vec{k_1} - \vec{k},l,n_4,s_2}  \right) \nonumber
 \\
 & \cross \delta[-n_3+n_4-n_1+n_2] \nonumber \\
 &+ \psi_{0,n_1,s_1}^{*}  \psi_{0,n_2,s_2} a_{s_1,s_2} \left( g_{\vec{k_1} + \vec{k},l,n_3,s_1} g^{*}_{\vec{k_1} + \vec{k},l,n_4,s_2} + f^{*}_{\vec{k_1} - \vec{k},l,n_3,s_1} f_{\vec{k_1} - \vec{k},l,n_4,s_2}  \right) \delta[-n_3+n_4-n_1+n_2] 
 \label{int_lhy_stripe_aux_2}
\end{align}
where $s_1$ and $s_2$ are spinor component indexes, $\mu$ is the chemical potential, $n$ is the density and we have expanded the Bogoliubov amplitudes of Eq.~\ref{qfluc_stripe} in Bloch waves~\cite{martone18}.
\begin{align}
 \vec{f}_{\vec{k_1} + \vec{k},l}(\vec{k},\vec{r}) = \frac{1}{\sqrt{V}} e^{i (\vec{k}_1 + \vec{k}) \vec{r}} \sum_{n \in \mathbb{Z}} \vec{f}_{\vec{k_1} + \vec{k},l,n} e^{2 i n k_1 x} 
 \label{f_p_stripe}
 \\
 \vec{f}^{*}_{\vec{k_1} - \vec{k},l}(\vec{k},\vec{r}) = \frac{1}{\sqrt{V}} e^{i (\vec{k}_1 + \vec{k}) \vec{r}} \sum_{n \in \mathbb{Z}}  \vec{f}^{*}_{\vec{k_1} - \vec{k},l,n} e^{2 i n k_1 x} 
  \label{f_m_stripe}
 \\
 \vec{g}_{\vec{k_1} - \vec{k},l}(\vec{k},\vec{r}) = \frac{1}{\sqrt{V}} e^{i (\vec{k}_1 - \vec{k}) \vec{r}} \sum_{n \in \mathbb{Z}}  \vec{g}_{\vec{k_1} - \vec{k},l,n} e^{2 i n k_1 x} 
  \label{g_p_stripe}
 \\
 \vec{g}^{*}_{\vec{k_1} + \vec{k},l}(\vec{k},\vec{r}) = \frac{1}{\sqrt{V}} e^{i (\vec{k}_1 - \vec{k}) \vec{r}} \sum_{n \in \mathbb{Z}}  \vec{g}^{*}_{\vec{k_1} + \vec{k},l,n} e^{2 i n k_1 x} 
 \label{g_m_stripe}
\end{align}
with $\vec{f}_{\vec{k},l,n} = (f_{\vec{k},l,n,s=+1} \text{   } f_{\vec{k},l,n,s=-1})^{\tau}$. The same holds for $\vec{g}_{\vec{k},l,n}$. The terms $\psi_{0,n,s}$ correspond to the expansion in Bloch waves of the condensate wave function (see the main text). The integration region in Eqs.~\ref{int_lhy_stripe_aux_0}-~\ref{int_lhy_stripe_aux_2} is $k_{\perp} = \sqrt{k_y^2 + k_z^2} \in [0,\infty)$, $0 < k_x < k_1$, with $k_1$ the ground state momentum. The sum indexes $\{ l,n_1,n_2,n_3,n_4 \}$ range from $-\infty$, $+\infty$. In practice, we introduce cut-off values in both operations and restrict the calculation to $0 < k_{\perp} = \sqrt{k_y^2 + k_z^2} < k_{\perp,\text{max}}$ and $-N_{c} < l,n_1,n_2,n_3,n_4 < N_{c}-1$. The integration volume is then $ V_{I} = \pi k_{\perp,\text{max}}^2 \times 4 N_{c} k_1$, a cylinder of radius $k_{\perp,\text{max}}$ in the $\{ k_y,k_z \}$ plane and height $4 N_{c} k_1$ in the $k_x$ axis, centered at the origin. The Bloch amplitudes fulfill the normalization condition~\cite{martone18}:
\begin{align}
\sum_{n=-N_{c}}^{n=N_{c}-1} \vec{f}^{\tau}_{k_1+\vec{k},l,n} \vec{f}_{k_1+\vec{k},l,n} - \left( \vec{g}^{\tau}_{k_1+\vec{k},l,n} \vec{g}_{k_1+\vec{k},l,n} \right) &= 1
\\
\sum_{n=-N_{c}}^{n=N_{c}-1} \vec{f}^{\tau}_{k_1-\vec{k},l,n} \vec{f}_{k_1-\vec{k},l,n} - \left( \vec{g}^{\tau}_{k_1-\vec{k},l,n} \vec{g}_{k_1-\vec{k},l,n} \right) &= -1
\end{align}
We define $N_x$ and $N_{\perp}$ as the number of points in the $x$ and radial axes, respectively. Looking at Eq.~\ref{int_lhy_stripe}, one notices that the integral scales as $\order{N_x N_{\perp} N_{c}^4}$, while typically, $N_x \sim \order{10^2}$, $N_{\perp} \sim \order{10^3}$ and the calculation becomes too expensive in computational cost terms. In order to make it feasible, we introduce two approximations. The first one involves the number of momentum components of the condensate wave function (i.e. indexes $n_1$ and $n_2$ in Eq.~\ref{int_lhy_stripe}). According to Ref.~\cite{martone18} and to our Simulated Annealing calculations, the absolute value of the Bloch wave amplitudes in the condensate wave function decreases very rapidly with the momentum index, $n$. Therefore, we denote by $N_{c,0}$ the number of momentum components of the condensate wave function included in the computation of the LHY integral and fix its value. In this way, the LHY integral scales as $\order{N_x N_{\perp} N_{c,0}^2 N_{c}^2}$, with the integration volume remaining unchanged. In practice, no significant changes are seen in the results when $N_{c,0}>5$, so we set $N_{c,0}=5$.

\begin{figure}[t]
\centering
\includegraphics[width=0.9\linewidth]{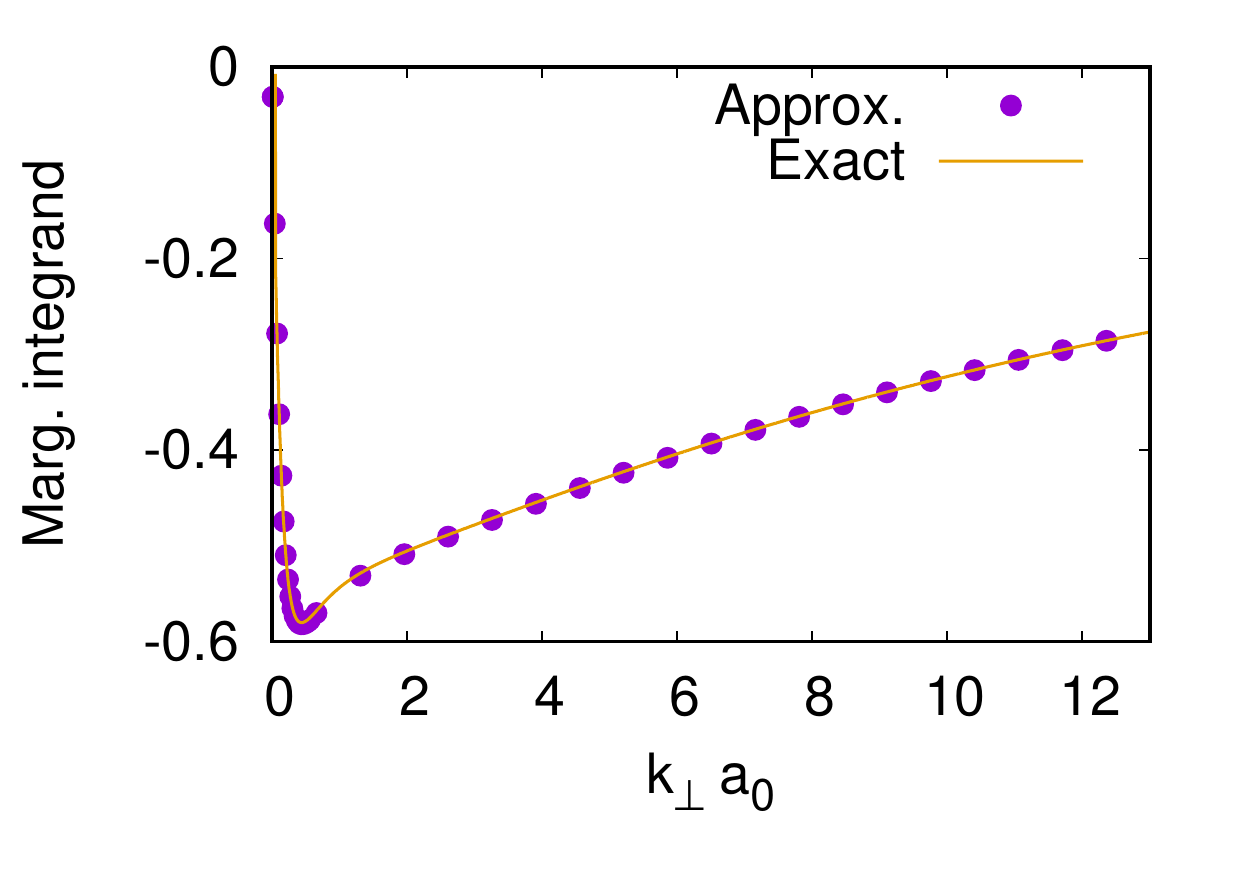}
\caption{Exact ($N_c=9$) and approximated ($N_c=9$, $N_{c,0}=5$) marginal integrands of the unregularized LHY energy per particle for $\Omega=2.8$, $a_{+1,+1} = a_{-1,-1} = 0.641982$, $\gamma = 0.4$, $n=3.7 \times 10^{-3}$, $N_x = 200$, $N_{\perp} = 2000$.} 
\label{fig_higher_momentum_approx}
\end{figure}

The computation cost can be furtherly reduced introducing a second approximation. It can be checked numerically that, as $k_{\perp}$ and $N_c$ increase, the integral $I_2$ is dominated by the contributions from the $f_{\vec{k_1}+ \vec{k},l,l,\pm 1}$ and $g_{\vec{k_1} - \vec{k},l,l,\pm 1}$ terms. Therefore, we retain the two dominant terms for every value of $l$ to the integral instead of performing the whole sum over $n_3$ and $n_4$. Additionally, we retain only the two first momentum modes of the condensate state when computing the integral $I_3$, since we have checked that these are the dominant contributions. These approximations reduce the scaling of the LHY integral on $N_c$ up to $\text{Max}\left\{ \order{N_x N_{\perp} N_{c} N_{c0}^2 },\text{ } \order{N_x N_{\perp} N_{c}^2 } \right\}$. We show in Fig.~\ref{fig_higher_momentum_approx} the marginal integrand of $E_{\text{LHY}}/N$ after integrating over the $x$-axis and performing the sums, for the exact case with $N_c=9$ and the approximated case with $N_c=9$, $N_{c,0} = 5$. As it can be seen from the Figure, both curves are in excellent agreement.

\begin{figure}[t]
\centering
\includegraphics[width=0.92\linewidth]{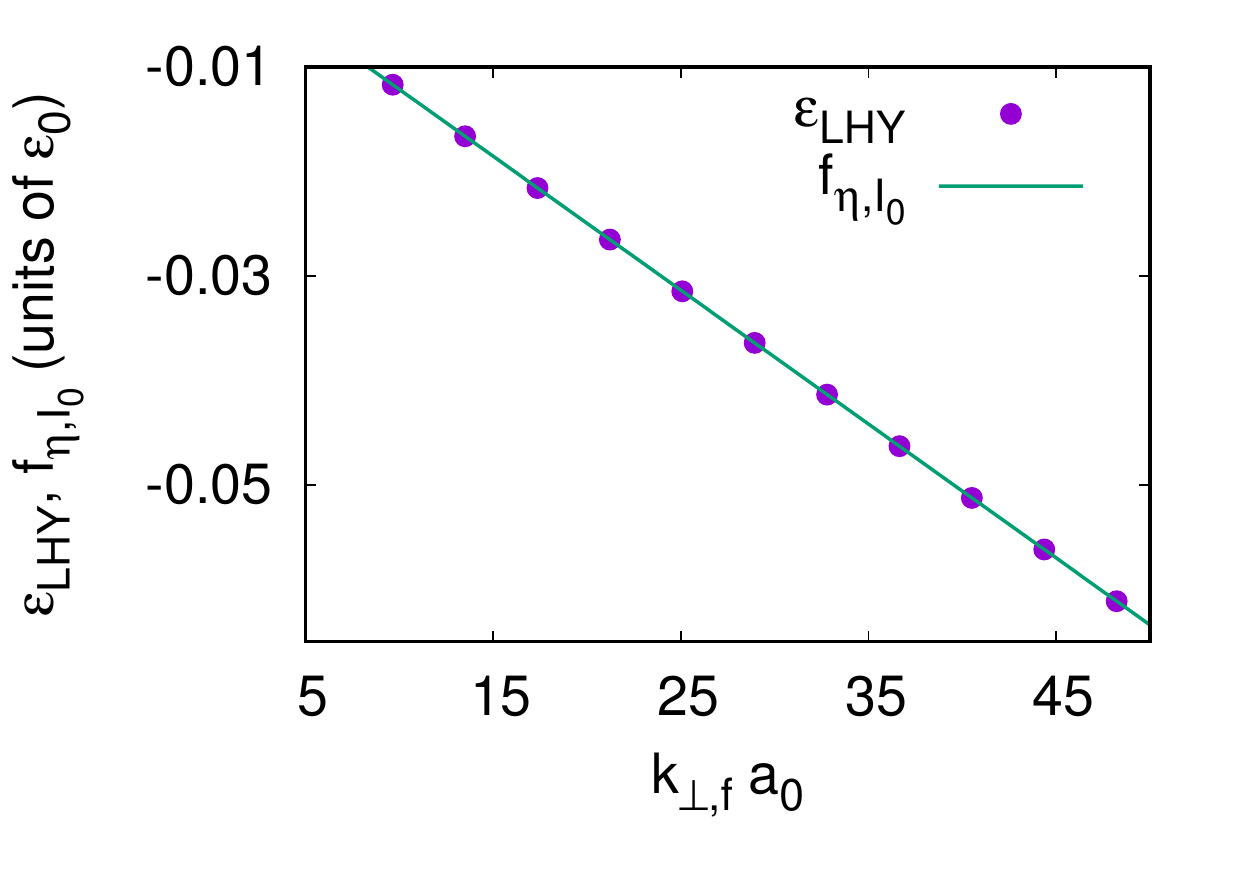}
\caption{$\epsilon_{\text{LHY}}$ (blue dots) and $f_{\eta, I_0}$ (green line) computed for different integration volumes $V_I$, with $V_I = \pi k_{\perp,\text{f}}^2 \times 2 k_{\perp,\text{f}}$, a cylinder of radius $k_{\perp,\text{f}}$ and height $2 k_{\perp,\text{f}}$, with $k_{\perp,\text{f}} = 2 N_c k_1$, $N_c \in [5,25]$. Other parameters are $\Omega=1.0$, $a_{+1,+1} = a_{-1,-1} = 0.2$, $\gamma = -21$, $n=3.11 \times 10^{-3}$, $N_x = 300$, $N_{\perp} = 3000$.} 
\label{fig_divergent_integral_fit}
\end{figure}

\begin{figure}[b]
\centering
\includegraphics[width=0.92\linewidth]{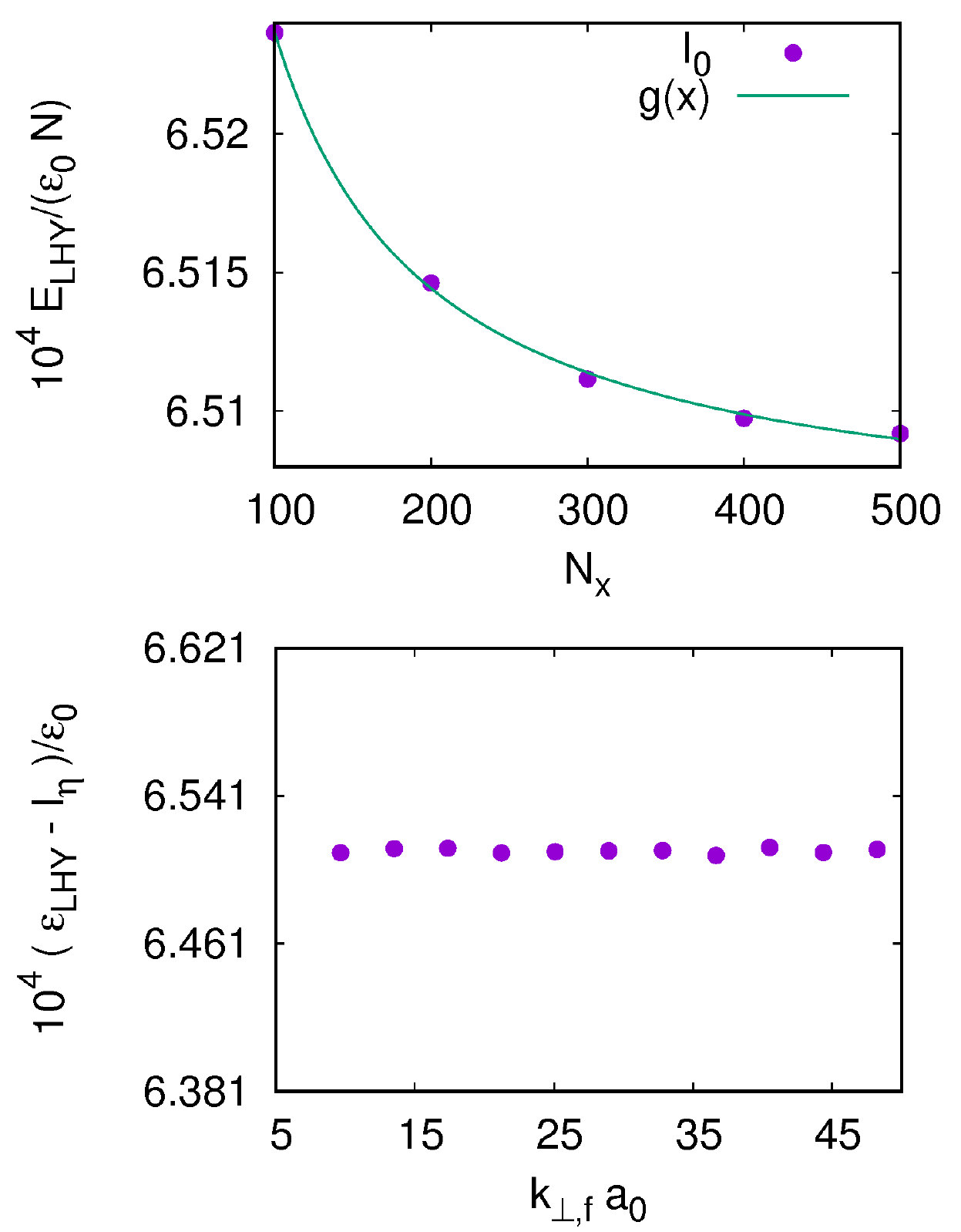}
\caption{Upper plot: $I_0(N_x)$ vs $N_x$. Lower plot: $\epsilon_{\text{LHY}} (V_I) - I_{\eta}(V_I)$ computed for different integration volumes $V_I$, with $V_I = \pi k_{\perp,\text{f}}^2 \times 2 k_{\perp,\text{f}}$, a cylinder of radius $k_{\perp,\text{f}}$ and height $2 k_{\perp,\text{f}}$. Other parameters are: $\Omega=1.0$, $a_{+1,+1} = a_{-1,-1} = 0.2$, $\gamma = -21$, $n=3.11 \times 10^{-3}$.} 
\label{fig_convergence_lhy}
\end{figure}  

\vspace{0.5cm}

\noindent
\large{\textbf{Regularization of the LHY integral}}

\vspace{0.5cm}

We define $\epsilon_{\text{LHY}} (V_I)$ as the integral of Eq.~\ref{int_lhy_stripe} over a finite integration volume $V_I$. As mentioned in the main text, the LHY integral for a Raman SOC stripe system is ultraviolet divergent (i.e. $\lim_{V_I \rightarrow \infty} \epsilon_{\text{LHY}}(V_I) = \infty$), and must be regularized. To identify the diverging behavior, we compute $\epsilon_{\text{LHY}} (V_I)$ over increasingly larger cylindrical volumes. These volumes are defined as $V_I^{(i)} = \pi (k^{(i)}_{\perp,\text{max}})^2 \times 4 N^{(i)}_{c} k_1$, with $k^{(i)}_{\perp,\text{max}} = 2 N^{(i)}_{c} k_1$, $N^{(i)}_{c} \in \mathbb{N}$. We find that $\epsilon_{\text{LHY}} (V_I)$ can be fitted to
\begin{equation}
 f_{\eta, I_0}(V_I) = \int_{V_I} \vec{dk} \text{ } \eta / k^2 + I_0 = I_{\eta}(V_I) + I_0
\end{equation}
with $\eta$ and $I_0$ fitting parameters, $k^2 = k_x^2 + k_y^2 + k_z^2$ and $I_{\eta}(V_I)$ given by:
\begin{equation}
 I_{\eta}(V_I) = 8 \pi \eta N_{c} k_1 \left( \frac{\pi}{4} + \frac{\log 2}{2} \right)
\end{equation}
where $\log$ indicates the natural logarithm. We show $\epsilon_{\text{LHY}} (V_I)$ and $f_{\eta, I_0}(V_I)$ as a function of the integration volume in Fig.~\ref{fig_divergent_integral_fit}. 
Therefore, the quantity $\lim_{V_I \rightarrow \infty} \epsilon_{\text{LHY}} (V_I) - I_{\eta}(V_I)$ is finite. Thus, the LHY energy per particle can be computed as:
\begin{equation}
 E_{\text{LHY}}/N = \lim_{V_I \rightarrow \infty} \left[ \epsilon_{\text{LHY}} (V_I) - I_{\eta}(V_I) \right] + I^{\text{reg.}}_{\eta}
\end{equation}
with $I^{\text{reg}}_{\eta}$ the regularized $I_{\eta}(\infty)$ value, which we obtain applying Dimensional Regularization~\cite{salasnich16,leibbrandt75}. In this scheme, the regularized integral of a polynomial identically vanishes~\cite{salasnich16}, which implies $I^{\text{reg}}_{\eta}(V_I = \infty) = 0$. Thus, the regularized LHY integral is given by:
\begin{equation}
 E_{\text{LHY}}/N =  \lim_{V_I \rightarrow \infty} \left[ \epsilon_{\text{LHY}} (V_I) - I_{\eta}(V_I) \right] = I_0
\end{equation}

\vspace{0.5cm}

\noindent
\large{\textbf{Convergence of the regularized LHY integral}}

\vspace{0.5cm}

Ideally, the regularized LHY integral should be computed for $N_c \rightarrow \infty$, $N_x \rightarrow \infty$, $N_{\perp} \rightarrow \infty$. However, in practice, the values $N_c$, $N_x$ and $N_{\perp}$ used in the calculations are finite. In order to approach the asymptotic limit, the regularized LHY integral is computed for different values $N_c \in [n_{c,0}, n_{c,1}]$ and for different number of points, $N_x \in [n_{x,0}, n_{x,1}]$, with $N_{\perp} = 10 N_x$. For each value of $N_c$, the cylindrical integration volume is $V_I = \pi (2 N_{c} k_1)^2 \times 4 N_{c} k_1$, analogously to the previous Section. For each fixed number of points, the fitting described in the previous Section is carried out, resulting on a function $I_0(N_x)$. We then extrapolate $I_0(N_x)$ to $N_x \rightarrow \infty$ using a function of the form $g(N_x) = a + b/N_x^l$ and take the extrapolation, $I_0(N_x \rightarrow \infty)$, as the final result. The range of $N_c$ is chosen such that the quantity $\epsilon_{\text{LHY}} (V_I) - I_{\eta}(V_I)$ does not depend on $N_c$, meaning that the asymptotic limit has been reached. We show in Fig.~\ref{fig_convergence_lhy}, $I_0(N_x)$ as a function of the number of points $N_x$ for $\Omega=1.0$, $a_{+1,+1} = a_{-1,-1} = 0.2$, $\gamma = -21$, $n=3.11 \times 10^{-3}$, and the quantity $\epsilon_{\text{LHY}} (V_I) - I_{\eta}(V_I)$ as a function of the number of modes $N_c$, for $N_x=300$. As it can be seen from the Figure, $\epsilon_{\text{LHY}} (V_I) - I_{\eta}(V_I)$ shows no significant dependence on $N_c$. The extrapolation of $I_0(N_x)$ to $N_x \rightarrow \infty$ yields the final result $E_{\text{LHY}}/N = I_0(N_x \rightarrow \infty) = 6.505 \times 10^{-4}$. In practice, one can just perform the calculations for two values of $N_c$ and one for $N_x$ such that $I(N_x) \simeq I(N_x \rightarrow \infty)$. As an example, setting $N_c = 5, 7$ and $N_x = 600$ one obtains $E_{\text{LHY}}/N = I_0(N_x = 600) = 6.494 \times 10^{-4} \simeq I_0(N_x \rightarrow \infty)$

\end{document}